\documentclass[1p]{elsarticle}
\usepackage{amsfonts,amsmath,amssymb}

\begin{document}
\title{Horava-Lifshitz cosmology, entropic interpretation and quark-hadron phase transition}
\author[sbu]{F. Kheyri}
\ead{F$\_$Kheyri@sbu.ac.ir}

\author[sbu]{ M. Khodadi}
\ead{M.Khodadi@sbu.ac.ir}

\author[sbu]{Hamid Reza Sepangi}
\ead{hr-sepangi@sbu.ac.ir}

\address[sbu]{Department of Physics, Shahid Beheshti University, G. C.,
Evin,Tehran 19839, Iran}

\begin{abstract}
Based on the assumptions of the standard model of cosmology, a phase transition associated with chiral symmetry breaking after the electroweak transition has occurred at approximately $10\mu$ seconds after the Big Bang to convert a plasma of free quarks and gluons into hadrons. We consider such a phase transition in the context of a deformed Horava-Lifshitz cosmology. The Friedmann equation  for the deformed Horava-Lifshitz universe is obtained using the entropic interpretation of gravity, proposed by Verlinde. We investigate the effects of the parameter $\omega$ appearing in the theory on the evolution of the physical quantities relevant to a description of the early universe, namely, the energy density and temperature before, during and after the phase transition. Finally, we study the cross-over phase transition in both high and low temperature regions in view of the recent lattice QCD simulations data.
\end{abstract}
\maketitle
\section{Introduction}\label{sec1}
The concept of gravity as a fundamental interaction has been under scrutiny ever since  the seminal work of Sakharov \cite{27}, the so called induced gravity, suggesting that the spacetime background emerges as a mean field approximation to the underlying microscopic degrees of freedom \cite{22}. Some years later, the four laws of black hole thermodynamics  \cite{23} were discovered and the entropy law was gradually extended to explain more general systems, known today as the holographic principle \cite{18}. The strongest evidence of the holographic principle was provided by the $AdS/CFT$ correspondence which states that all  information about a gravitational system in a region of spacetime is contained in its boundary \cite{17}. The relation between gravity and thermodynamics was studied in \cite{21} where it was argued that  the Einstein field equations is a consequence of the relation between the horizon area, entropy and the Clausius relation $\delta Q = Tds$, where $\delta Q$ and $T$ are the energy flux and Unruh temperature seen by an accelerated observer just inside the horizon. A number of different approaches were subsequently followed where the emphasis was on the close connection between Einstein gravity and thermodynamics \cite{19}, \cite{25} and \cite{28}. Along this line, a suggestion has  been made in the recent past  by Verlinde \cite{6} to the effect that gravity can be considered as an entropic force caused by changes in the information related to the positions of material bodies. This idea implies that gravity is not a fundamental force and, somehow,  has its roots laid out in the work presented in \cite{27} more than forty years ago.
In addition, to understand the universe at its earliest moments, a quantum theory of gravity should be constructed for which all attempts have been fruitless so far. Therefore the idea of an entropic force is interesting since it could be instrumental in explaining  the origin of gravity and its unification with  quantum theory. The spacetime can be considered as an information device built up of holographic screens or surfaces over which the the system's information can be stored. The screens or surfaces behave like stretched horizons in  black hole physics. The information on such screens is described by the entropy and  is encoded in terms of the number of bits which, naturally,  are proportional to the area of the screen \cite{23}. In \cite{60} the author, using the Verlinde's idea describes the Coulomb�s law in the framework of the entropic force. In \cite{61}, the authors argue that the entropic formulation provides new insights into the quantum nature of the inertia and  gravity. They use the entropic postulate to determine the quantum uncertainty in the law of inertia and the law of gravity in both non-relativistic and  relativistic cases.  As discussed in \cite{62} the entropic gravity cannot explain observations related to the gravitationally-bound quantum states of neutrons in the GRANIT experiment. Later, a careful analysis \cite{63} showed that such analysis is not correct \cite{64}.

It is a well known fact that the biggest challenge towards a theory of quantum gravity is the question of renormalizability whose lack, in effect, would imply loss of theoretical control and predictability at high energies.  Many attempts have been made in the past decades to address this question, but to no avail. However, a recent  proposition made by Horava suggests that a Lifshitz type anisotropic scaling at high energies can be used to evade this difficulty \cite{1,2}. This theory, known as Horava-Lifshitz gravity, was motivated by a scalar field model proposed by Lifshitz \cite{3} to explain the quantum critical phenomena in condensed matter physics where the corresponding action possesses $z =2$ scale-invariance  where $z$ represents the dynamical critical exponent given by
\begin{equation}
t\rightarrow\ell^{z}t,\quad x^{i}\rightarrow \ell x^{i}.
\end{equation}
The theory thus obtained is non-relativistic in the $UV$ limit and is hoped to be $UV$ finite. Horava has argued that it is superficially renormalizable based on power counting and may lead to Einstein's general relativity in the IR region. Therefore, it may be regarded as a UV complete candidate for general relativity. Although the gravity action of the Horava theory with the detailed balance condition recovers  GR in the IR limit,  its solutions may not do so. For example the authors in \cite{16} have studied thermodynamics of black holes in the deformed HL gravity with a dynamical coupling constant $\lambda$ and showed that for  $\lambda=1$, the black hole behaves like a Reissner-Norstrom black hole which is different from the Schwarzschild black hole solution obtained in Einstein gravity. However, in \cite{30} the authors studied a particular limit of the theory which admits flat Minkowski vacuum and discussed the quadratic fluctuations around it. They showed that by adding a term proportional to the spatial three-curvature $^3\!R$ to the action, the IR modification can be recovered which allows for the flat minkowski vacuum as a solution. In this way, the spherically symmetric, asymptotically flat black hole solution is constructed in GR. We note that the IR vacuum of Horava theory is an anti de Sitter (AdS) spacetime. In order to achieve a more desirable IR behavior without discarding the simplicity provided by the detailed balance condition in the UV limit, a non-trivial test of the new gravity theory in a FRW universe was studied by introducing the $\mu^{4}$ $^3\!R$ term in the second reference in \cite{30}. By taking the $\Lambda_{w}\rightarrow 0$ limit and considering an IR modification which breaks ``softly'' the detailed balance condition in the original Horava model, the IR modified Horava gravity seems to become consistent with the current observational data.

The main purpose of this paper is devoted to the study of QCD phase transition in the context of deformed HL gravity from the point of view of the entropic force, introduced by Verlinde to study the early universe. The standard model of cosmology suggests that as the early universe expanded and cooled, it underwent a series of symmetry-breaking phase transitions, causing topological defects to form. It is the study of such phase transitions that would pave the way for a better understanding of the evolution of the early universe, characterized by the existence of a quark-gluon plasma undergoing a phase transition \cite{32}. During the evolution of the very early universe there have been at least two phase transitions. The electroweak theory predicts that at about $100GeV$ there was a transition from a symmetric high temperature phase with massless gauge bosons to the Higgs phase, where the $SU(2)\times U(1)$ gauge symmetry is spontaneously broken and all the masses in the model are generated. Also, quantum chromodynamics (QCD) predicts that at about $200 MeV$ there was a transition from a quark-gluon plasma to a plasma of light hadrons, which is thought to have occurred early in the history of the universe. The taking place of this phase transition is dated  back to a few microseconds after the Big Bang for which the universe had a mean density of the same order as nuclear matter $(\rho\sim10^{15}\frac{gr}{cm^{3}})$.  In what follows we focus attention on possible scenarios which might have occurred to allow such a phase transition to come to the fore. We generally follow the discussion presented in \cite{31} and \cite{33} which puts the quark-gluon phase transition in a cosmologically transparent perspective.

The study of a phase transition in the early universe requires a knowledge of the equation of state
(EoS) which plays an essential role in the hot, strongly interacting matter and in the hydrodynamic
description of heavy ion collisions \cite{19, 20, 6}. Generally, for a description
of the collective flow in heavy ion collisions, model theories have been constructed with an EoS
representing a first order phase transition. However, lattice QCD calculations suggest that the
transition to the de-confined phase is that of a crossover. The EoS is not only of theoretical
interest, but is directly applicable to the dynamics of the quark-gluon plasma (QGP), whether
in the context of interpreting the results of heavy-ion experiments or modeling the behavior
of hot, dense matter in the early universe \cite{38}.The possibility of a phase transition
in the gas of a quark-gluon bag was demonstrated for the first time in \cite{34}. The
mechanism responsible for a first-order phase transition has been elegantly discussed in
\cite{33}. Attempts to calculate the EoS on the lattice have been made over the last $20$
years \cite{41}. As is well known, calculating  the equation of state of QCD is one of the central
goals of lattice simulations at finite temperature \cite{26}.  Lattice QCD is a fundamental tool which
makes a systematic study of the non-perturbative regime of the QCD equation of state possible.
For a comparison of lattice calculations of the EoS with continuum perturbation theory or
models like the MIT bag EoS, it is necessary to remove the lattice artifacts.
This approach using   supercomputers \cite{35} has led to the calculation of the QCD equation of state with two light
quarks with a heavier strange quark on a $(N_{\tau}=6\times32^{3})$ size lattice 
\cite{45,48, 26,42,44,47,43,50,52}.
\section{Friedmann equation in deformed Horava-Lifshitz gravity in the context of entropic force}\label{sec2}
In this section, we  study the Friedmann equation in deformed HL gravity in the framework of the entropic force. Let us start with a (3+1)-dimensional FRW universe which is described by the metric
\begin{equation}\label{e2-1}
ds^{2}=-dt^{2}+a(t)^{2}\left[\frac{dr^{2}}{1-kr^{2}}+r^{2}d \Omega_{2}^{2}\right],
\end{equation}
where $k$ denotes the spatial curvature constant with $k=-1,\,0,\,1$ corresponding to a closed, flat and open universe, respectively with $d\Omega_{2}^{2}$ representing the two-sphere. Introducing $\tilde{r}= a(t)r$, the metric (\ref{e2-1}) becomes
\begin{equation}\label{e2-2}
ds^{2}=h_{ab}dx^{a}dx^{b}+\tilde{r}^{2}d\Omega_{2}^{2}.
\end{equation}
with $x_{0}=0$, $x_{1}=r$ and $h_{ab}= \mbox{diag}\left(-1,\frac{a^{2}}{1-kr^{2}}\right)$.
A marginally trapped surface with vanishing expansion for a dynamical apparent horizon is determined by the relation $h_{ab}\partial_{a}\tilde{r}\partial_{b}\tilde{r}=0$, implying that the vector $\nabla\tilde{r}$ is null or degenerate on the apparent horizon surface \cite{24}. It is now a matter of a simple calculation to show that the dynamical apparent horizon, $\tilde{r}_{A}$, is given by
\begin{equation}\label{e2-3}
\tilde{r}_{A}=\frac{c}{\sqrt{H^{2}+\frac{k}{a^{2}}}},
\end{equation}
where $H=\frac{\dot{a}}{a}$ is the Hubble parameter and $c$ is the speed of light.
Now, suppose that the energy-momentum tensor $T_{\mu\nu}$ of the matter in the universe has the form of a perfect fluid $T_{\mu\nu}=(\rho+p)u_{\mu}u_{\nu}+p g_{\mu\nu}$, with $\quad u_{\mu}=-\delta^{t}_{\mu}$ where $\rho$, $p$ and $u_{\mu}$ are the energy density, isotropic pressure and four-velocity of the cosmological fluid, respectively. The matter conservation law,  $\nabla_{\mu}T^{\mu\nu}=0$, leads to
\begin{equation}\label{e2-4}
\dot{\rho}+3H(\rho+p)=0,
\end{equation}

In an interesting work \cite{11} it has been shown that  the Friedmann equation can be derived from the first law of thermodynamics, see  \cite{12, 13, 14} for further details. It is thus of interest to obtain the modified Friedmann equation in deformed HL gravity in the context of the entropic force. To begin we note that the temperature associated with apparent horizon can be defined as $T=\frac{K}{2\pi}$, where $K$ is the surface gravity given by $K=\frac{1}{\sqrt{-h}}\partial_{\mu}(\sqrt{-h}h^{\mu\nu}\partial_{\nu}\tilde{r})$. It can subsequently be shown that the temperature of the apparent horizon can be written as \cite{29}
\begin{equation}\label{e2-9}
T=\frac{\hbar c}{2\pi k_{B}\tilde{r}_{A}}.
\end{equation}
One of the quantum corrections applied to the horizon area law of a black hole is the logarithmic correction that arises from loop quantum gravity (LQG) due to thermal equilibrium fluctuations and quantum fluctuations. As a result, the entropy is written as \cite{120}
\begin{equation}
S=\frac{k_BAc^{3}}{4G\hbar}+\frac{k_B\alpha}{4}\ln\left(\frac{Ac^{3}}{G\hbar}\right),
\end{equation}
where the geometric parameter $\alpha$ is a dimensionless constant of order unity whose exact value is still a puzzle in LQG. It is clear that if $\alpha=0$, then the entropy expression reduces to Schwarzschild's entropy, $S=\frac{A}{4\ell^{2}_{p}}$ where $\ell^{2}_{p}=\frac{G\hbar}{c^{3}}$ is the Planck length. The geometric parameter $\alpha$ has a direct relation to gravity and therefore by applying the logarithmic corrections to the entropy of black holes in the deformed HL gravity one has \cite{16,36}
\begin{equation}\label{e2-10}
S=\frac{k_BAc^{3}}{4G\hbar}+\frac{k_B\pi}{\omega}\ln\left(\frac{Ac^{3}}{G\hbar}\right).
\end{equation}
One notes that in HL theories $\alpha$ is proportional to $\omega^{-1}$ where $\omega$ is a dimensionless constant parameter of the theory. One also notes that for $\omega\rightarrow\infty$ we discover the entropy as obtained in Einstein gravity. 

Verlinde \cite{6} made an interesting suggestion that gravity may be regarded as an entropic force caused by  changes in the information associated with the positions of material bodies. Such a strong assumption together with the holographic principle and the equipartition law of energy, enabled Verlinde to obtain the second law of Newtonian mechanics and the Newtonian law of gravitation. Verlinde postulated that when a test particle with $mass$ approaches a holographic screen from a distance $\Delta x$, the change of entropy on the holographic screen is
\begin{equation}\label{e10}
\Delta S=2\pi k_{B}\frac{\Delta x}{l_{c}}
\end{equation}
where $l_{c}=\frac{\hbar}{mc}$ is the Compton wavelength. Thus, the Compton wavelength defines the units of the entropy change. From this relation we realize that when a test particle which has  mass $m$ and is located at the distance $\Delta x$ from the holographic surface moves to the immediate vicinity of the surface, it causes a change of  entropy $\Delta S$ of the holographic screen. In effect, this means that the particle is embedded into the surface and becomes  part of the information and energy on the screen.
According to Verlinde's argument, gravity arises from the entropic gradient $\triangle S$ over the spatial region $\triangle x$ via the first law of thermodynamics
\begin{equation}\label{e2-16}
F=T\frac{\triangle S}{\triangle x},
\end{equation}
where $T$ is the temperature of the screen and $F$ is the so-called entropic force. Originally, the starting point of Verlinde's derivation was an assumption that the holographic screen forms a  sphere with radius $r$. It is also known that  a sphere with area $A$ acts as a storage device for information, that is
\begin{equation}\label{e2-17}
N=\frac{Ac^{3}}{G\hbar}.
\end{equation}
Verlinde then argued that the energy $E$ of a system is the rest energy $E=Mc^{2}$ of the mass inside the sphere and that the energy is divided evenly over the bits, $N$
\begin{equation}\label{e2-18}
E =\frac{1}{2}N k_{B}T.
\end{equation}
It is now seen that if we insert the relation $E=Mc^{2}$ in equation (\ref{e2-18}) with thermodynamical equilibrium in  mind, that is, $T=T_{U}$ and use the Unruh effect \cite{65}  given by $T_{U}= \frac{\hbar a}{2\pi k_{B}}$ where $a$ is the proper acceleration of the observer and also use the fact that the area of a two-sphere with radius $r$ is $A=4\pi r^{2}$, we find
\begin{equation}\label{e2-19}
F=G\frac{mM}{r^{2}},
\end{equation}
which is Newton's universal law of gravitation. A very similar derivation of equation\, (\ref{e2-19}) was also discovered by Padmanabhan \cite{40}. Using the entropy of a black hole which is a quarter of the area of the event horizon  \cite{39},   one may write
\begin{equation}\label{e2-20}
N=\frac{4S}{k_{B}}.
\end{equation}
Plugging equation\,(\ref{e2-10}) in equation\,(\ref{e2-20}), we obtain the total number of bits for the HL gravity in the form
\begin{equation}\label{e2-21}
N=\frac{Ac^{3}}{G\hbar}+\frac{4\pi}{\omega}\ln\left(\frac{Ac^{3}}{G\hbar}\right).
\end{equation}

The Friedmann equation may also be reproduced from this entropic force method. In the context of the entropic force, the energy $E$ for a thermodynamic system is defined by the number of bits on the holographic screen by the equipartition law $E =\frac{1}{2} Nk_{B}T$ with $T$  the temperature of the holographic screen.
Here, we choose the apparent horizon to be the holographic screen. So the temperature $T$ is that of the apparent horizon. Therefore $dE$ from the equipartition law becomes
\begin{equation}\label{e2-22}
dE =\frac{ 1}{2} Nk_{B}dT+ \frac{1}{2} k_{B}T dN
   =\frac{8\pi G\hbar+A\omega c^{3}-4\pi G\hbar\ln(\frac{Ac^{3}}{G\hbar})}{4A^{3/2}\sqrt{\pi}G\hbar\omega} c dA.
\end{equation}
Also, the amount of energy crossing the apparent horizon during the time interval $dt$ is defined as \cite{11}
 \begin{equation}\label{e2-15}
 dE=-A(\rho+p)H\tilde{r}_{A} dt.
 \end{equation}
From equations\,(\ref{e2-22}) and (\ref{e2-15}), we have
\begin{equation}\label{e2-23}
\left[8\pi G\hbar+A\omega c^{3}-4\pi G\hbar\ln\left(\frac{Ac^{3}}{G\hbar}\right)\right]d\tilde{r}_{A}
=16\pi^{2}c G\omega\tilde{r}_{A}^{5}(\rho+p)Hdt.
\end{equation}
After some algebra we find
\begin{equation}\label{e2-26}
\frac{G\hbar}{2\omega c^{5}}\left(H^{2}+\frac{k}{a^{2}}\right)^{2}\left[4-2\ln\left(\frac{4\pi c^{5}}{G\hbar\left(H^{2}+\frac{k}{a^{2}}\right)}\right)\right]+\left(H^{2}+\frac{k}{a^{2}}\right)=\frac{8\pi\rho}{3}.
\end{equation}
Equation\, (\ref{e2-26}) is the modified Friedmann equation from the entropic force which is more complicated compared to the standard Friedmann equation.
The above equation can rewritten in a more manageable form, see the second reference in \cite{29}
\begin{equation}\label{e2-27}
\frac{1}{2\omega}\left(H^{2}+\frac{k}{a^{2}}\right)^{2}+\left(H^{2}+\frac{k}{a^{2}}\right)=\frac{8\pi\rho}{3}.
\end{equation}
As expected, setting $\omega\rightarrow\infty$, we see that the first term in equation\,(\ref{e2-26}) vanishes and the standard Friedmann equation is recovered. Here $\rho$ is the total energy density, $k$ is the spatial curvature of the universe, $H=\frac{\dot{a}}{a}$ is the Hubble parameter and $a$ is the scale factor. Equation\,(\ref{e2-27}) is a second degree equation in  $(H^{2}+\frac{k}{a^{2}})$ with a solution
\begin{equation}\label{e2-28}
\left(H^{2}+\frac{k}{a^{2}}\right)=\omega\left(-1+\sqrt{1+\frac{16\pi}{3\omega}\rho}\right).
\end{equation}
where we have set $\hbar=c=G=k_{B}=1$. From equation\,(\ref{e2-28})  one notes that according to the
entropic interpretation of the force of gravity, the term $H^{2}$ is proportional to $\sqrt{\rho}$.
\section{Quark-hadron phase transition}\label{sec3}
When the Hubble radius was around $10km$ and the universe was about $t \sim 10^{-5}s$ old, the last phase transition predicted by the standard model of particle physics took place at the QCD scale $T \sim 200$ MeV.  In order to study the quark-hadron phase transition in the deformed Horava-Lifshitz gravity theory we need  the equation of state of the matter, in both quark and hadron states. We shall use the simplest possible form of an EoS with a first order phase transition, that of a bag EoS.  The smallness of baryon entropy ratio  in the early universe  allows us to restrict ourselves to the case where the chemical potential is very small, $\mu\ll T$ which means  that we can determine the EoS by ignoring the effects of chemical potential  \cite{46}. Let us start from the equation of state of matter in the quark phase 
\begin{equation}\label{e3-1}
\rho_{q}=3a_{q}T^{4}+V(T),\quad p_{q}=a_{q}T^{4}-V(T),
\end{equation}
where $a_{q}=(\pi^{2}/90)g_{q}$ with $g_{q}=g_{Q}+(g_{b}+\frac{7}{8}g_{f})$ and $g_{Q}=16+(21/2)N_{F}$ is the intrinsic statistical weights for the quark-gloun plasma and $g_{b},g_{f}$ are the statistical weights of bosons and fermions, respectively. At the epoch of the quark-hadron phase transition photons $(g_b=2)$, electrons $(g_f=4)$, muons$(g_f=4)$ and neutrinos $(g_{f}=6)$ yield $g_{b}+7/8 g_{f}=14.25$. Here $N_{f}$ is the number of relativistic quark flavors, two at lower temperature corresponding to the u, d quarks and three at higher temperatures where the strange quark becomes relativistic. One may then write $N_{F}=2 $, $ g_{q}=51.25$.  The self-interaction potential $V(T)$ is defined as \cite{49}
\begin{equation}\label{e3-2}
V(T)=B+\gamma_{T}T^{2}-\alpha_{T}T^{4},
\end{equation}
where $B$ is the QCD vacuum energy or bag pressure constant. Results obtained in low energy hadron spectroscopy, heavy ion collisions and phenomenological fits of light hadron properties give $B^{1/4}\in(100-200)MeV$, $\alpha_{T}=7\pi^{2}/20$ and $\gamma_{T}=m_{s}^{2}/4$ \cite{51} with $m_{s}$ being the mass of the strange quark, considered in the range $m_s \in(60 - 200)MeV$). In the case where the temperature effects are negligible, the equation of state in the quark phase takes the form of the MIT bag model equation of state, $p_{q}=\frac{\rho_{q}-4B}{3}$ in the hadron phase and we take the cosmological fluid with energy density $\rho_{h}$ and pressure $p_{h}$ as an ideal gas of massless pions and nucleons obeying Maxwell-Boltzmann statistics. The equation of state can now be approximated by
\begin{equation}\label{e3-3}
p_{h}(t)=\frac{\rho_{h}(T)}{3}=a_{\pi}T^{4},
\end{equation}
where $a_{\pi}=(\pi^{2}/90)g_{h}$ and $g_{h}=g_{H}+(g_{b}+\frac{7}{8}g_{f})$. Taking $g_{H}\approx3$ as the intrinsic statistical weights for the soup of hadrons, we have
$g_{h}=3+14.25=17.25$. During quark-hadron phase transition the temperature is equal to the critical temperature $T_{c}$ which is defined by the condition $p_{q}(T_{c})=p_{h}(T_{c})$, and is given by \cite{46}
 \begin{equation}\label{e3-4}
T_{c}=\sqrt{\frac{\gamma_{T}+\sqrt{\gamma_{T}^{2}+4(a_{q}+\alpha_{T}-a_{\pi})B}}{2(a_{q}+\alpha_{T}-a_{\pi})}}.
\end{equation}
If we take $m_{s}=200 MeV$ and $B^{1/4} = 200 MeV$, the transition temperature is of the order $T_{c}=125MeV$. One should note that since the phase transition is assumed to be of first order, all the physical quantities like energy density, pressure and entropy exhibit discontinuities across the critical curve. Ignoring temperature effects in $V(T)$ and taking  $\alpha_{T}=\gamma_{T}\approx0$  in the above equation we obtain the well known relation between  $T_{c}$ and $B$, that is  $B=\frac{(g_{q}-g_{h})\pi^{2}T_{c}^{4}}{90}$ \cite{46}.

Now, two thermodynamic quantities are of importance in the study of the early universe. The first is the enthalpy defined as  $W=U+pV$  where $U = E$ is the total energy flow through the apparent horizon which leads to  $W=\frac{4}{3}(\rho+p)\pi \tilde{r}_{A}^{3}$, from which the heat capacity of the universe enveloped by the apparent horizon at constant pressure  can be obtained
\begin{equation}\label{e3-5}
C_{p}=\left(\frac{\partial W}{\partial T}\right)_{p}=\left(\frac{\partial W}{\partial\tilde{r}_{A}}\frac{\partial\tilde{r}_{A}}{\partial T}\right)_{p}.
\end{equation}
The other thermodynamic quantity is the entropy, defined as $S=V\frac{dp}{dT}$ (chemical potential is zero). Since in the expanding  universe the change of the horizon radius is positive and that of the horizon temperature is negative we have $\frac{\partial\tilde{r}_{A}}{\partial T}<0$ thereupon $C_{p}<0$ which points to the fact that the system (universe) is losing energy in the form of heat.
\section{Dynamical consequences of deformed horava-lifshitz universe during quark-hadron phase transition}\label{sec4}
Let us now study the phase transition described above. To this end we obtain the physically important quantities in the quark-hadron phase transition in deformed HL theory like the energy density $\rho$, temperature $T$ and scale factor $a$. These parameters are derived from the deformed Friedmann equation (\ref{e2-27}), conservation equation (\ref{e2-4}) and equations of state (\ref{e3-1}), (\ref{e3-2}) and (\ref{e3-3}). We also focus attention on a spatially flat universe for which $k=0$.

\begin{figure}
\includegraphics[width=8cm]{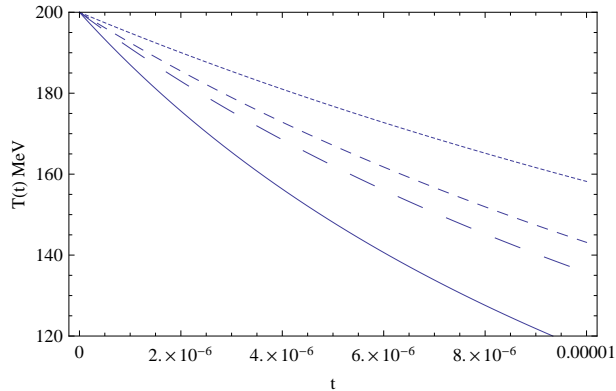}
\centering\caption{The behavior of temperature of the quark fluid as a function of $t$ for different values of $\omega$; $\omega= 10^{6}$, dotted curve, $\omega=5\times 10^{6}$, short-dashed curve, $\omega= 10^{7}$ long-dashed curve and $\omega=2\times 10^{7}$, solid curve. We have taken $ B^{\frac{1}{4}}=200MeV$.
}{\label{fig.1} \small}
\end{figure}

For $T > T_{C}$, before the phase transition, the universe is in  pure quark phase and $H$ is obtained from the conservation equation (\ref{e2-4})
\begin{equation}\label{e4-1}
H=-\frac{\dot{\rho}}{3(\rho+p)},
\end{equation}
where using the deformed Friedmann equation (\ref{e2-28}) we have
\begin{equation}\label{e4-2}
\dot{\rho}=-3(\rho+p)\left[\omega\left(-1+\sqrt{1+\frac{16\pi\rho}{3\omega}}\right)\right]^{1/2}.
\end{equation}
Finally,  substitution of equations of state of the quark matter (\ref{e3-1}) and self-interaction potential term (\ref{e3-2}) leads to
\begin{equation}\label{e4-3}
\frac{dT}{dt}=-\frac{12a_{q}T^{4}}{(12a_{q}-\alpha_{q})T^{3}+2\gamma_{T} T}
\left[\omega\left(-1+\sqrt{1+\frac{16\pi\rho}{3\omega}\left(3a_{q}T^{4}+\gamma_{T}T^{2}-\alpha_{T}T^{4}+B\right)}\right)\right]^{1/2}.
\end{equation}
Equation (\ref{e4-3}) may be solved numerically and the result is presented in figure \ref{fig.1} which shows the behavior of temperature as a function of the time $t$ in a deform HL world filled with quark matter for different values of $\omega$.

\begin{figure}
\includegraphics[width=8cm]{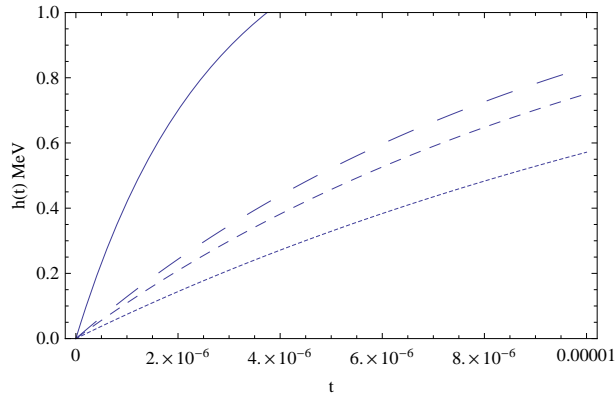}
\centering\caption{The behavior of $h(t)$ as a function of time $t$ for different values of $\omega$; $\omega= 10^{6}$, dotted curve) $\omega=5\times 10^{6}$, short-dashed curve, $\omega= 10^{7}$, long-dashed curve, $\omega=2\times 10^{7}$, solid curve}{\label{fig.3} \small}
\end{figure}

For $T = T_{C}$, during the phase transition, the temperature and the pressure are constant. In this case quantities like the entropy $ S = sa^{3}$ and  enthalpy $W =(\rho + p)a^{3}$ are conserved. For later convenience, following first reference in \cite{46}, we replace $\rho(t )$ by $h(t)$, so that the volume fraction of matter in the hadron phase is given by
\begin{equation}\label{e4-9}
\rho(t)=\rho_{H}(t)+\rho_{Q}(1-h(t))=\rho_{Q}(1+mh(t)),
\end{equation}
where $m=\frac{\rho_{H}}{\rho_{Q}}-1$. The beginning of the phase transition is characterized by $h(t_{c})= 0$ where $t_{c}$ is the time  and $\rho(t_{c}) \equiv\rho_{Q}$, while the end of the transition is characterized by $h(t_{h})=1$ with $t_{h}$ being the time signaling the end and corresponding to $\rho(t_{h})\equiv\rho_{h}$. For $t>t_{h}$
the universe enters into the hadronic phase. For the phase transition temperature of $T_{c}=125MeV$ we have $\rho_{Q}=5\times10^{9}MeV^{4}$ and $\rho_{H}=1.38\times10^{9}MeV^{4}$, respectively. For the same value of the temperature, the value of the pressure of the cosmological fluid during the phase transition is $p_{c}=4.6\times10^{8}MeV^{4}$. Now, substituting equation (\ref{e4-9}) in equation (\ref{e4-2}) leads to
\begin{equation}\label{e4-10}
\frac{dh}{dt}=-\frac{3(1+rh(t))}{r}\left[\omega\left(-1+\sqrt{1+\frac{16\pi}{3\omega}\rho_{q}(1+mh(t))}\right)\right]^{1/2},
\end{equation}
where $r=(\rho_{H}-\rho_{Q})/(\rho_{Q}+p_{c})$. Figure \ref{fig.3} shows variation of the hadron fraction $h(t)$ as a function of $t$ for different values of $\omega$.

\begin{figure}
\includegraphics[width=8cm]{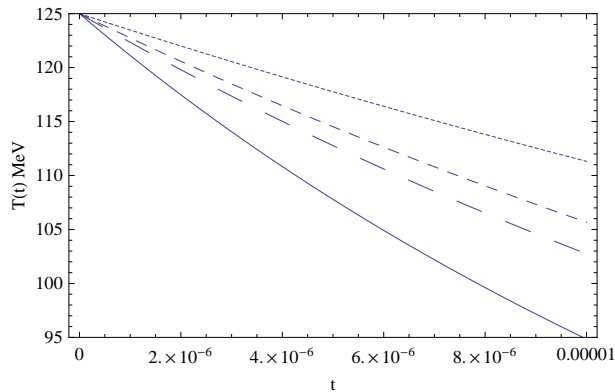}
\centering\caption{The behavior of $T(t)$ in a hadron phase as a function of time $t$ for different values of $\omega$; $\omega= 10^{6}$, (dotted curve, $\omega=5\times 10^{6}$, short-dashed curve, $\omega= 10^{7}$, long-dashed curve, $\omega=2\times 10^{7}$, solid curve. We have taken $T_{c}=125MeV$.
}{\label{fig.4} \small}
\end{figure}

For $T<T_{C}$, after the phase transition, the energy density of the pure hadronic matter is in form of equation\,(\ref{e3-3}). Substituting this equation in (\ref{e4-2}) leads to the time variation of temperature of the universe in the hadronic phase
\begin{equation}\label{e4-11}
\frac{dT}{dt}=-T\left[\omega\left(-1+\sqrt{1+\frac{16\pi}{\omega}a_{\pi}T^{4}}\right)\right]^{1/2}.
\end{equation}
Variation of temperature of the hadronic fluid filled deformed HL universe as a function of $t$ for different values of $\omega$ is represented in figure \ref{fig.4}.

\section{Lattice QCD phase transition}\label{sec5}
The importance of the equation of state in the understanding of thermal properties of any thermodynamic system is fundamental. One of the pivotal goals of non-pertubative studies of QCD on the lattice is  the understanding of the bulk thermodynamics of QCD \cite{48}.  Within the context of perturbation theory and the Hadronic Resonance Gas model (HRG) and at high and low temperature the equation of state can be determined \cite{58, 52}. However, in the transition region from the low temperature hadronic regime to the high temperature quark gluon plasma, one has to rely on a genuine non-perturbative approach, lattice regularized QCD, to study the non-perturbative properties of strongly interacting matter. Lattice studies of bulk thermodynamics observables, pressure and energy density, are given in terms of differences of dimension $4$ operators. These differences are particularly difficult to evaluate because both terms being subtracted contain the pressure or energy density of the vacuum, an unphysical quantity that is approximately $1/(aT)^{4}$ larger than the sought-after difference. Thus Numerical signals rapidly decrease with the fourth power of the lattice spacing, $a$, when one attempts to approach the continuum limit at a fixed  temperature $T$. For this reason, improved actions will allow one to perform calculations on rather coarse lattices with relatively small lattice discretization errors. The early calculations of bulk thermodynamics with standard staggered and Wilson fermion \cite{57} discretization schemes have shown that at high temperature bulk, thermodynamic observables are particularly sensitive to lattice discretization errors. In order to minimize discretization errors at high temperature, improved staggered fermion P4 and asqtad actions have been used to study the QCD equation of state \cite{55}.

Lets now discuss the EoS  obtained in lattice QCD in the framework of the crossover phase transition. In the first reference in \cite{55}, the authors have analyzed the hot QCD collaboration data by two different improved staggered fermion action, P4 and asqtad,  together with physical $s$ quark mass and the up and down quark with masses larger than the physical mass $\frac{m_{s}}{m_{u,d}}=10$, have determined a transition region in the range $T\in(185-195)MeV$. However, in \cite{56} the authors, using  staggered stout action which take the up, down  and strange quarks with the physical mass $\frac{m_{s}}{m_{u,d}}=28$, have shown that the transition region stays in the range $T\in(150-170)MeV$. Recent results for different physical observables with EoS at $N_{t}=6,8,10$ can be found in \cite{53}. In this work we use the EoS given by the HRG model in low temperature region together with lattice QCD data in high temperature region which have been combined with a simple parametrization \cite{52}. The trace anomaly $\Theta(T)=\rho-3p$ crucial in lattice determination of the equation of state since in lattice QCD the calculation of the pressure, energy density and entropy density usually done through the computation of the trace anomaly. The pressure difference at temperatures $T$ and $T_{low}$ can be expressed as the integral of the trace anomaly
\begin{equation}\label{e5-1}
\frac{p(T)}{T^{4}}-\frac{p(T_{low})}{T_{low}^{4}}=\int_{T_{low}}^{T}\frac{dT'}{T'^{5}}(\rho-3p).
\end{equation}
If the lower integration limit is taken to be sufficiently small, the pressure at this temperature becomes an exponential suppression and can be neglected. This procedure is known as the integral method \cite{54}. The energy density then becomes $\rho(T)=\Theta(T)+3p(T)$. Finite temperature lattice calculations are usually performed at fixed temporal extent $N_{t}$ and the temperature is varied by varying the lattice spacing $a $ with $T = 1/(N_{t}a)$ . Hence, the lattice spacing becomes smaller as the temperature  increases and thus the trace anomaly can be accurately calculated in the high temperature region. In consequence, to construct a realistic equation of state we use the lattice data for the trace anomaly in the high temperature region, $T > 250MeV$ \cite{52}.
Lattices with asqtad and P4 actions within the HRG model are in good agreement with lattice QCD data for temperature region below $180 MeV$. Therefore, in the HRG model the trace anomaly can be  calculated with enough accuracy in the low temperature region, $T\leq180 MeV$. It can be shown that \cite{55, 48} in the high temperature region the trace anomaly may be parameterized by an inverse polynomial in the form
\begin{equation}\label{e5-2}
\frac{(\rho-3p)}{T^{4}}=\frac{d_{2}}{T^{2}}+\frac{d_{4}}{T^{4}}+\frac{c_{1}}{T^{n_{1}}}+\frac{c_{2}}{T^{n_{2}}}.
\end{equation}
 \begin{table}
\caption{The values of parameters for different fits to the trace anomaly \cite{52}.}
\begin{tabular}{|c|c|c|c| c| c| c| c|}
  \hline
  & $d_{2}(GeV)$ &$d_{4}(GeV^{4})$ &$c_{1}(GeV^{n_{1}})$&$c_{2}(GeV^{n_{2}})$&$n_{1}$&$n_{2}$&$T_{low}(MeV)$\\ \hline
  $s95p$ &$0.2660$ &$2.403\times10^{-3} $&$-2.809\times10^{-7} $&$6.073\times10^{-23}$&$ 10$&$ 30$&$183.8$\\
$s95n$&$0.2654$&$6.563\times10^{-3}$&$-4.370\times10^{-5}$&$5.774\times10^{-6}$&$8$&$9$&$171.8$\\
$s90f$&$0.2495$&$1.355\times10^{-2}$&$-3.237\times10^{-3}$&$1.439\times10^{-14}$&$5$&$18$&$170.0$\\
  \hline
  \end{tabular}
\end{table}

This form does not have the right asymptotic behavior in the high temperature region where we expect $(\rho-3p)/T^{4} = g^{4}(T) 1/ \ln^{2}(T/\Lambda_{QCD})$,  but works well at the temperature range of interest. The parametrization of the trace anomaly and therefore QCD equation of state calculated using such requirements are labeled by $s95p-v1$, $s95n-v1$ and $s90f-v1$, as shown in table 1. The labels ``s95'' and ``s90'' point to the fraction of the ideal entropy density  at $T=800MeV$ (95\% and 90\% respectively) whereas the labels $p$, $n$ and $f$ point to a specific consideration of the peak of the trace anomaly and its matching to the HRG. The values of the parameters $T_{low}$, $d_{2}$, $d_{4}$, $c_{1}$, $c_{2}$, $n_{1}$ and $n_{2}$ in each case are shown in table 1.
Inserting equation (\ref{e5-2}) in equation (\ref{e5-1}) for  high temperature, the pressure and energy density are given by
\begin{eqnarray}\label{e5-3}
p(T)=-\frac{1}{4}d_{4}-\frac{d_{2}}{2}T^{2}
+(\alpha_{i}+\beta_{i})T^{4}-
\frac{c_{1}}{n_{1}}T^{4-n_{1}}-\frac{c_{2}}{n_{2}}{T^{4-n_{2}}},
\end{eqnarray}
\begin{equation}\label{e5-4}
\rho(T)=3(\alpha_{i}+\beta_{i})T^{4}-\frac{1}{2}d_{2}T^{2}+
\frac{1}{4}d_{4}+\frac{c_{1}}{n_{1}}(n_{1}-3)T^{4-n_{1}}
+\frac{c_{2}}{n_{2}}(n_{2}-3)T^{4-n_{2}},
\end{equation}
where $i=1,2,3$ correspond to different parameterizations present in the above table, so that
\begin{equation}\label{e5-5}
\alpha_{i}=-(\frac{1}{4}d_{4}+\frac{d_{2}}{2}T_{low}^{2}+ \frac{c_{1}}{n_{1}}T_{low}^{4-n_{1}}+\frac{c_{2}}{n_{2}}{T_{low}^{4-n_{2}}}),
\hspace{3mm} \beta_{i}=p_{0}/T_{0}^{4},
\end{equation}
where $\alpha_{i}$ refer to the value of integration in the lower limit. Next, by substituting (\ref{e5-3}) and (\ref{e5-4}) in equation (\ref{e4-2}), the basic equation describing the change of temperature in the high temperature region becomes
\begin{eqnarray}\label{e17}
\frac{dT}{dt}=&-&\frac{(-d_{2}T^{2}+4(\alpha_{i}+\beta_{i})T^{4}+\frac{c_{1}}{n_{1}}(n_{1}-4)T^{4-n_{1}}+\frac{c_{2}}{n_{2}}(n_{2}-4)
T^{4-n_{2}})}{12(\alpha_{i}+\beta_{i})T^{3}-d_{2}T+\frac{c_{1}}{n_{1}}(n_{1}-3)(4-n_{1})T^{3-n_{1}}+\frac{c_{2}}{n_{2}}
(n_{2}-3)(4-n_{2})T^{3-n_{2}}} \hspace{3cm}\nonumber\\
&\times&\left[\omega\left(-1+\sqrt{1+\frac{16\pi}{3\omega}\rho(T)}\right)\right]^{1/2},\vspace{0.5cm}
\end{eqnarray}
\begin{figure}
\includegraphics[width=8cm]{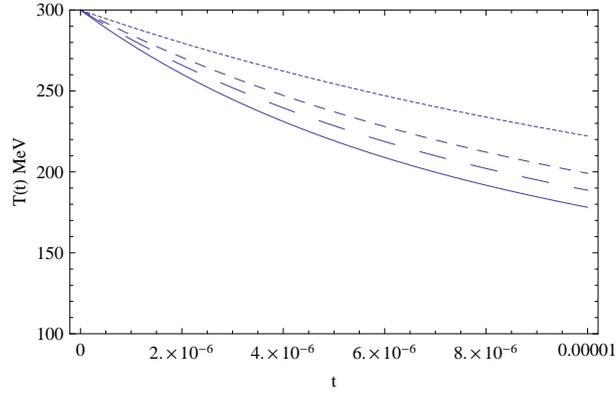}
\centering\caption{Variation of temperature in the high temperature region as a function of $t$ for the first parametrization corresponding to $T_{0}=170.0 MeV$. The behavior of temperature is shown for different values of  $\omega$; $\omega= 10^{6}$, dotted curve, $\omega=5\times 10^{6}$, short-dashed curve, $\omega= 10^{7}$, long-dashed curve and $\omega=2\times 10^{7}$, solid curve {\label{fig.5} \small}}
\end{figure}

\begin{figure}
\includegraphics[width=8cm]{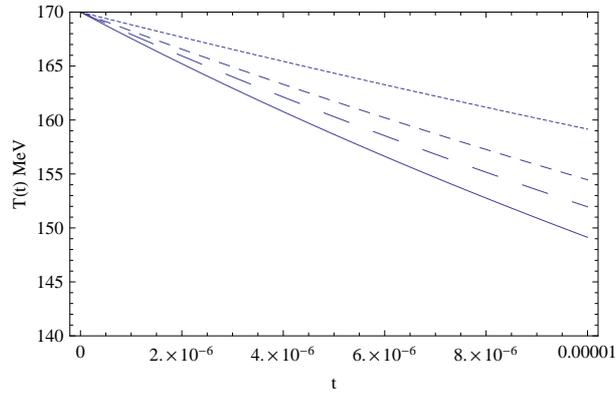}
\centering\caption{Variation of temperature in the low temperature region as a function of $t$ corresponding to $T_{0}=183.8 MeV$. The behavior of temperature is shown for different values of  $\omega$; $\omega= 10^{6}$, dotted curve, $\omega=5\times 10^{6}$, short-dashed curve, $\omega= 10^{7}$, long-dashed curve and $\omega=2\times 10^{7}$, solid curve. {\label{fig.8} \small}}
\end{figure}

\begin{figure}
\includegraphics[width=8cm]{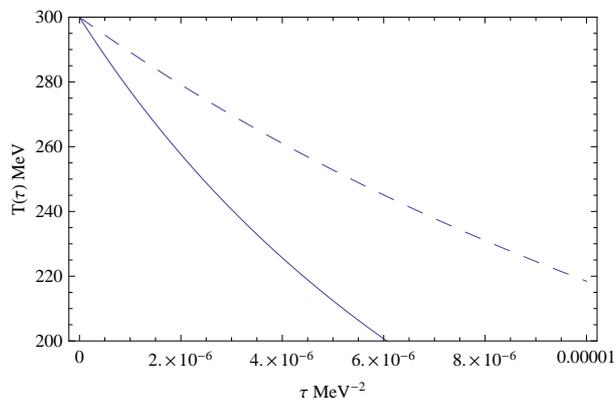}
\centering\caption{Comparison of the effective temperature in the high temperature region with the effective temperature of quark-gluon plasma for $\omega= 10^{6}$. The dashed curve shows the high temperature region and the solid curve shows the quark phase. {\label{fig.9} \small}}
\end{figure}
\begin{figure}
\includegraphics[width=8cm]{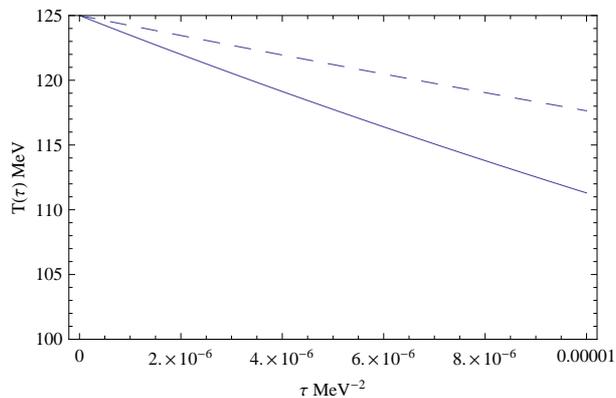}
\centering\caption{Comparison of effective temperature in the low temperature region with the effective temperature of hadronic phase for $\omega= 10^{6}$. The dashed curve shows the low  temperature region and the solid curve shows the hadronic phase. {\label{fig.10} \small}}
\end{figure}
\noindent where $\rho(T)$  above refers to equation (\ref{e5-4}).
Apart from lattice QCD there are other approaches to the low temperature equation of state. In the context of the Hadronic Resonance Gas model, the confinement phase of QCD is treated as a non-interacting gas of fermions and bosons \cite{50}.  The HRG model gives a decent description of thermodynamic quantities in the transition region. The HRG result for the trace anomaly can also be parameterized by the simple form
\begin{equation}\label{e5-7}
\frac{(\rho-3p)}{T^{4}}=a_{1}T+a_{2}T^{3}+a_{3}T^{4}+a_{4}T^{10},
\end{equation}
where $a_{1}=4.654GeV^{-1}, a_{2}=-879GeV^{-3}, a_{3}=8081GeV^{-4}$ and $a_{4}=-7039000GeV^{-10}$. Inserting this equation  in equation (\ref{e5-1}) for the low temperature, the pressure and energy density can be found
\begin{equation}\label{e5-8}
p(T)=(\alpha_{0}+\beta_{0})T^{4}+a_{1}T^{5}+\frac{1}{3}a_{2}T^{7}+\frac{1}{4}a_{3}T^{8}+\frac{1}{10}a_{4}T^{14},
\end{equation}
\begin{equation}\label{e5-9}
\rho(T)=3(\alpha_{0}+\beta_{0})T^{4}+4a_{1}T^{5}+2a_{2}T^{7}+\frac{7}{4}a_{3}T^{8}+\frac{13}{10}a_{4}T^{14},
\end{equation}
with
\begin{equation}\label{e5-10}
\alpha_{0}=-(a_{1}T_{low}+\frac{a_{2}}{3}T_{low}^{3}+ \frac{a_{3}}{4}T_{low}^{4}+\frac{a_{4}}{10}T_{low}^{10})\hspace{1.2cm}
\beta_{0}=p_{low}/T_{low}^{4}=0.1661,\quad T_{low}=0.07 Gev.
\end{equation}
In a similar fashion, substitution of the EoS equation (\ref{e5-8}) and (\ref{e5-9}) in equation (\ref{e4-2}) yields the basic equation describing the change of temperature in the high temperature region
\begin{equation}
\frac{dT}{dt}=-\frac{(12\alpha_{0}+6\beta_{0})T^{4}+15a_{1}T^{5}+7a_{2}T^{7}+6a_{3}T^{8}+
\frac{6}{5}a_{4}T^{14}}{(12\alpha_{0}+4\beta_{0})T^{3}+20a_{1}T^{4}
+14a_{2}T^{6}+14a_{3}T^{7}+\frac{21}{5}a_{4}T^{3}}\hspace{9cm}\nonumber\\
\times\left[\omega\left(-1+\sqrt{1+\frac{16\pi}{3\omega}\rho(T)}\right)\right]^{1/2},\hspace{6cm}
\end{equation}
where $\rho(T)$ above refers to equation (\ref{e5-9}).
Figure (\ref{fig.8}) shows such a variation of temperature in the low temperature region.
In addition, we have compared the effective temperature of the first order phase transition with that of the cross-over, which is shown in  figures (\ref{fig.9}) and (\ref{fig.10}). As can be seen, the effective temperature of the lattice data which correspond to a cross-over phase transition is higher than the effective temperature of a first order phase transition.

\section{Conclusions and remarks}
In this work, we have discussed the quark-hadron phase transition in the context of a deformed Horava-Lifshitz gravity scenario  within an effective model of quantum chromodynamics. By using the Friedmann equations based on entropic interpretation of gravity proposed by Verlinde, we studied the evolution of the physical quantities relevant to the physical description of the early universe; the energy density and
temperature, before, during, and after the phase transition. We find that for
different values of $\omega$, phase transition occurs and the results show that increasing  this
parameter decreases the effective temperature of the quark-gluon plasma and of the hadronic fluid,
and also increases the hadronic fraction in the mixed phase.
In the last section we have considered the quark-hadron transition in the context of a
smooth crossover regime at high $(T>250  Mev)$ and low $(T < 180Mev)$ temperatures. For both
regions, we used the equation of state obtained by the hot ``QCD'' collaboration. The results
which are based on lattice data for a parametrization of high temperature region
and a parametrization for low temperature region represent the same effect on the variation of
temperature similar to the quark-gluon plasma and  hadronic phases. The generic behavior of the
temperature of the early universe in such a scenario is similar to that of a first order
phase transition, although the differences in the energy should be taken into account.
As mentioned above, the effective temperature of a first order phase transition is lower than that of a cross-over since the temperature falls off  faster in a first order phase transition. Therefore,  hadronization takes place more rapidly in a universe undergoing a first order phase transition.



\end{document}